# Geoweb 2.0 for Participatory Urban Design: Affordances and Critical Success Factors

Burak Pak and Johan Verbeke

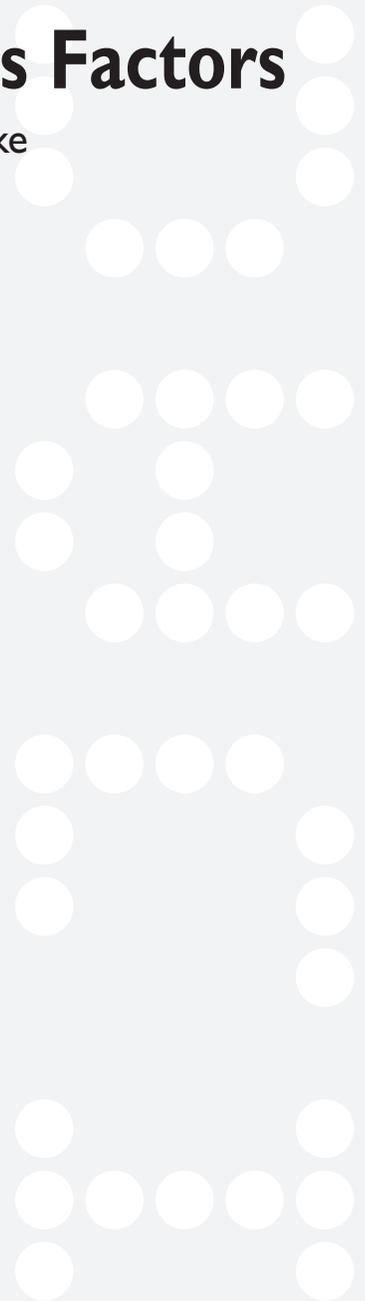

# Geoweb 2.0 for Participatory Urban Design: Affordances and Critical Success Factors

Burak Pak and Johan Verbeke

In this paper, we discuss the affordances of open-source Geoweb 2.0 platforms to support the participatory design of urban projects in real-world practices. We first introduce the two open-source platforms used in our study for testing purposes. Then, based on evidence from five different field studies we identify five affordances of these platforms: conversations on alternative urban projects, citizen consultation, design empowerment, design studio learning and design research. We elaborate on these in detail and identify a key set of success factors for the facilitation of better practices in the future.

# 1. INTRODUCTION

This paper is a comprehensive account of the findings from a postdoctoral research project between 2009 and 2013 which was supported by the Brussels Institute for the Encouragement of Scientific Research (INNOVIRIS) and KU Leuven Faculty of Architecture. The project aimed at developing and testing alternative strategies and tools for facilitating participatory urban design and learning processes. The motivations for the study were threefold:

- The need for enabling affordable participatory urban design tools and strategies
- Exploring the potentials of geographic and social Web 2.0 (Geoweb 2.0) technologies
- Understanding the use of alternative urban projects as a reflective resource

The first motivation was based on a common observation: traditional participation models rely on meetings held at a fixed time and place that are often ineffective and inefficient which severely limits the number of stakeholders taking part [1][2]. In such meetings, it is almost impossible to establish communication that is equal and accessible. It is also difficult to engage a high number of citizens into the early stages of the design process and take their ideas into account [3]. Too often the involvement of the participants is superficial and simply used to bring credibility to the design and decision-making process without really transforming these [4]. Consequently, the ordinary citizens and civil society are isolated from the design and decision-making process, leading to a shared sense of alienation and lack of trust in the society [5].

Second, as potential solutions to the issues addressed in our first motivation, Geoweb 2.0 platforms are well-positioned for facilitating dialogue and learning as well as communicative action [6][7]. "Web 2.0" term used in this context does not only refer to a jargon but it is also a *transforming prefix*. This prefix refers to a variety of constructivist strategies, tools and techniques that encourage and augment informed, creative and social interactions [8]. These strategies potentially enable production methods which can be considered as strong alternatives to the traditional, linear, hierarchical and centralized knowledge production methods.

During the last decade, these potentials have been demonstrated in a plethora of domains such as water management, soil management, ecosystem and forest management, habitat planning and native vegetation management, environmental planning [9], wind energy planning [10] as well as transportation and land use planning [11]. In these domains, fundamental changes are taking place in the way researchers deal with the issues of planning and governance. In the field of interdisciplinary computational modeling, problem-focused participatory mapping practices are transforming the existing analysis and decision networks [12]. Furthermore, there is a rising interest in the inclusion of social factors and involving human actors in scientific research by establishing a social-ecological system through coupling human and biophysical systems [13]. In contrast with the traditional modes of scientific production, the knowledge created through these constructivist methods are socially distributed, application oriented, transdisciplinary and subject to multiple accountabilities [14]. However, these methods don't necessarily describe a specific content to be used as context-setting element. In our case the content refers to *urban projects*. Hence, our third motivation involves the reframing of alternative urban projects as a public knowledge resource which can possibly stimulate civic engagement, dialogue and emergence of novel ideas [15].

Reflecting on the three motivations above, the following research questions will be addressed in this study:

- Can open-source Geoweb 2.0 platforms *afford* the deliberation and participatory design of alternative urban projects in real-life practices? If yes, to what extent?
- What are the key influence factors to be considered for the facilitation of better Geoweb 2.0 supported urban design practices?

In this context, we will start our paper by introducing two Geoweb 2.0 platforms used in our study (Section 2). Afterwards, we will reveal five affordances of these platforms providing evidence from our five field-studies and situate these in a global frame, making links with the existing body of literature and similar practices (Sections 2.1-2.5). Following this discussion, in Section 3, we will share the lessons learned from the field and identify a key set of factors for implementation and facilitation of better practices in the future.

## 2. AFFORDANCES OF TWO GEOWEB 2.0 PLATFORMS

In this paper, the term "affordance" is used to describe the "*action possibilities latent in the environment in relation to agents and their capabilities*" as introduced by Gibson [16]. According to this theory, the affordances of an environment are what it offers to the users, what it provides or furnishes. The "*environment*" in this study entails two open-source Geoweb 2.0 Platforms (P1 and P2) enhanced with several libraries and modules. We will start this section with a brief review of these platforms.

*The Geoweb 2.0 platforms used in this study*
The first platform (P1) is based on the MediaWiki content management software. MediaWiki uses an extensible lightweight wiki markup language and contains a variety of functionalities including rich content, an editing interface, search function, media library and an application-programming interface. We embedded the GoogleAPI in this system via "Google MediaWiki Extension" developed by Evan Miller [17]. The semantic mapping functionality was made available through "Semantic Maps Extension"

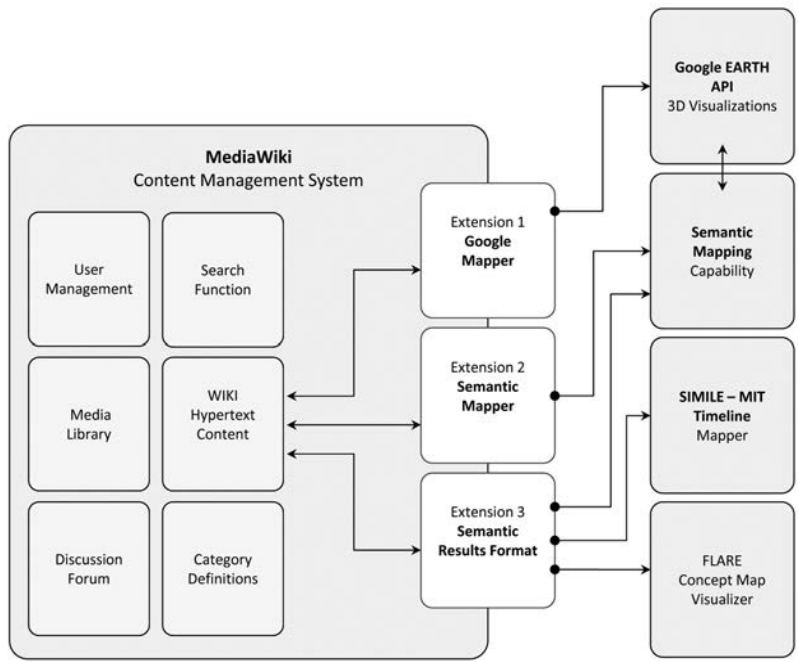

◄ Figure 1. Schematic Description of the Geoweb 2.0 Platform P1.

developed by de Dauw et al. [18]. The timelines and concept maps have been connected to related "SIMILE" and "FLARE" visualization libraries by Semantic Results Formats extension by Dengler et al. [19] (Figure 1).

MediaWiki was preferred as a backbone because it has successfully served to millions of contributors of Wikipedia as a robust platform. Moreover, several studies suggested it as a potential collaboration medium. For instance, Burrow and Burry [20] reported the effective use of Wikis as an internationally distributed design research network incorporating diverse forms of expertise. Chase et al. [21] introduced the "Wikitecture" concept as a decentralized method of open source co-production and tested the use of a Wiki to collaboratively develop a design competition entry.

The second platform (P2) used in this study is based on more than twenty open-source content management modules and other custom applications (Figure 2). It was constructed after the testing P1 with the contribution of real-life practitioners. Therefore it can be considered as a more developed and customized platform. In P2, Openlayers serves as the key library module for creating location-based information as well as complex geocoding and visualization. It provides the ability to connect to any popular mapping API available, including Google Maps, Bing Maps and OpenstreetMaps (Figure 2).

In this setup, jQuery and its user interface (UI) library provide abstractions for low-level interactions as well as advanced effects and themeable widgets. Geotaxonomy was used to attach geo information (latitude, longitude, bounding boxes, etc.) to taxonomy terms. Similar to the

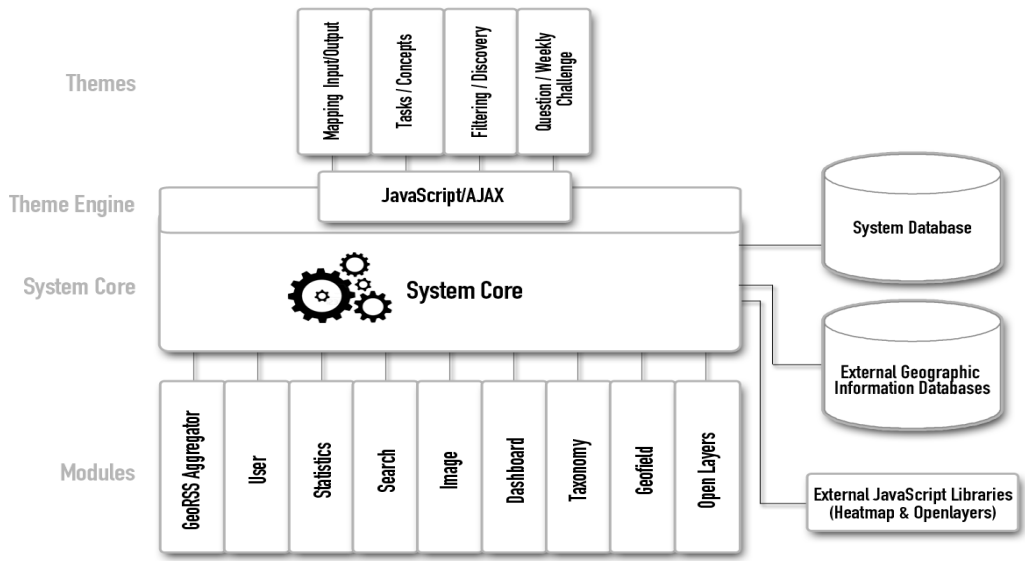

◄ Figure 2. Schematic Description of the Geoweb 2.0 Platform P2.

first platform, Flare library has been integrated into the system, this time through jQuery.

*Affordances and the Users*

Between 2009 and 2013, we have tested the two platforms introduced above in numerous scenarios and contexts [22][23][24]. Based on our findings from these real-world cases, observations and field experiences, we derived five action possibilities for the deliberation and participatory design of alternative urban projects in relation to specific users:

1. **Conversations on Alternative Urban Projects** – Users: Civil Society, Planners, Development Agencies
2. **Citizen Consultation** – Users: Youngsters, Non-profit environmental organizations and managers.
3. **Urban Design Empowerment** – Users: Non-profit planning organizations
4. **Learning in Urban Design Studio** – Users: Urban Design Students, Teachers, Experts, and Inhabitants.
5. **Urban Design Research** – Users: Urban Design Researchers, Inhabitants and Planners

In the next part of our paper, these *affordances* will be situated in a global frame and discussed with evidence from our real-life cases, making links with the existing body of literature and similar studies.

## 2.1. Affordance 1: Conversations on Alternative Urban Projects

Alternative Urban Projects (AUPs) are potential sources of knowledge because they include diagnoses of characteristic features and shortcomings of an urban situation and propose ideas for future development. Although it may not be possible to implement these proposals as a whole, the ideas and resources they accommodate can be used as a basis to develop new proposals [15].

AUPs simultaneously cover representations of the existing urban environment and imaginations of different realities which provide different frameworks for the discussion of the contemporary situation of the urban context. Initiating a sustainable conversation on these projects which have not been realized can lead to the creation of valuable knowledge. In this context, Geoweb 2.0 platforms can be framed as a potential interface between the decision-makers and the civil society (Figure 3).

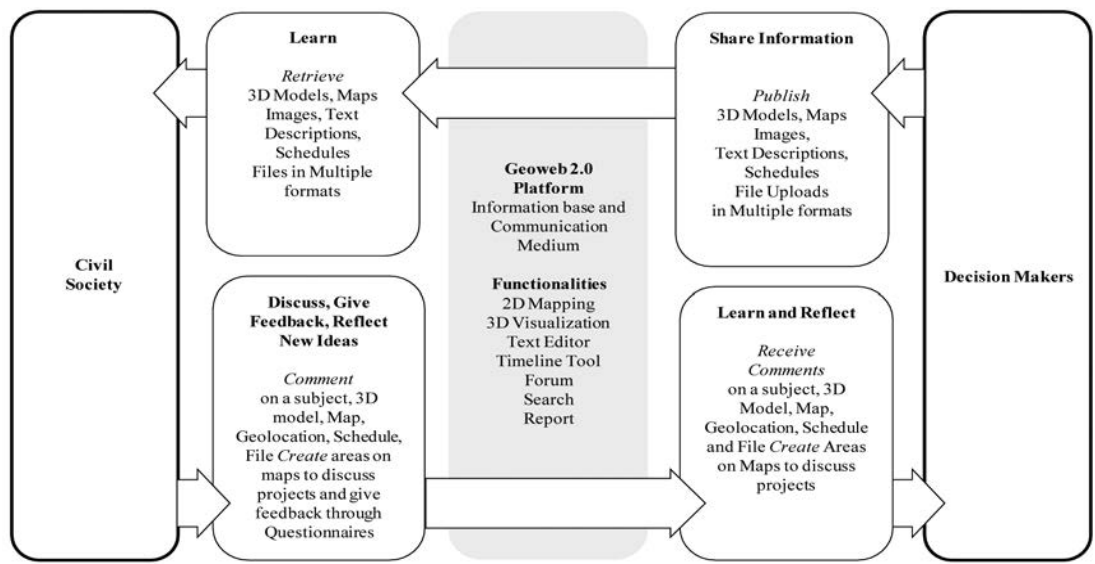

▲ Figure 3. Geoweb 2.0 as an interface between the decision-makers and the civil society for facilitating conversations on Alternative Urban Projects.

*Field Study 1: Conversations on Alternative Projects*

Motivated by the potentials referenced above, the authors constructed a use scenario with contributions of the Brussels Territorial Development Agency (ATO). The research method employed in this study involved analysis, synthesis and evaluation phases which runs on two parallel tracks: urban design and ICT. In the first phase, urban projects prepared for the two zones in Brussels (1990-2009) were analyzed. This analysis was accompanied by a review of available technologies and methods that support evaluation and deliberation of these projects. In the second phase, a synthesis of the findings from the analysis of the urban project were made. Furthermore, through numerous focus meetings with two NGOs (BRAL and Green Belgium) and the development agency ATO, a use scenario was developed. In this scenario, the Geoweb 2.0 platform was situated as an interface through which civil society and practitioners can learn, exchange ideas and shape the future strategies. Besides containing alternative urban projects, it aimed at initiating a dialogue by allowing professionals to publish information on their development projects. In this way, the authors intended to encourage the civil society to discuss, create ideas and give feedback in reflective manner.

Considering these observations, the use scenario and the feedback of the institutional actors (including a requirements analysis questionnaire), a conceptual design was developed. This design was a web application hybrid based on a combination of different representations, organized in two parts (Figure 4). As an initial study, the proposed platform P1 was used to implement the functions described in the conceptual design phase, focusing on two strategic zones in the city.

From this practice, we learned that geolocation can serve as the key integration tool for representation and discussion of the projects. The georeferenced definition of the strategies and spatial alternatives provided a potential framework for a structured and place-based dialogue. The actors were able to add comments as place marks and create a map of the debate on zones of interest.

Within the proposed framework it was also possible to divide a large body of text into individual place-based strategies and geolocate them on a map as interrelated place marks with explanations (Figure 4). Setting the goals of the study with the contribution of real actors ensured the synchronization of the platform with their needs. As a result it inspired the initiation of Bruplus, a novel Geoweb 2.0 platform covering an overview of all major urban projects in Brussels.

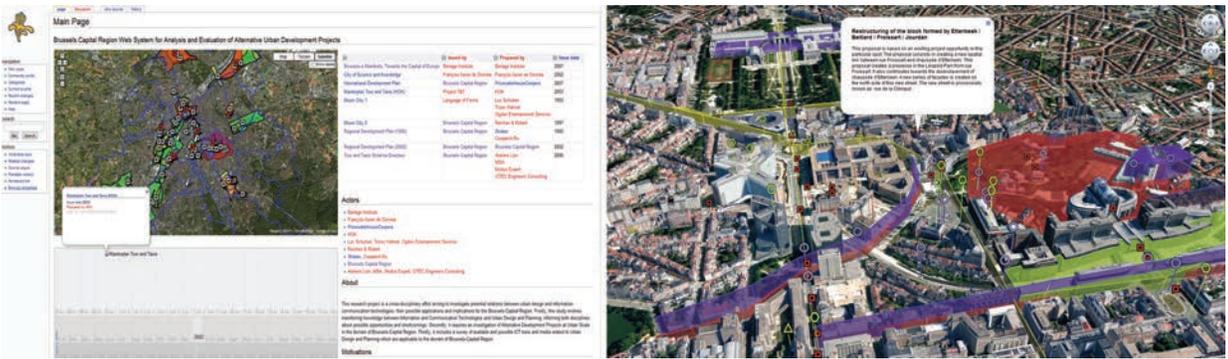

▼ Figure 4. Preliminary implementation of platform P1.

## 2.2. Affordance 2: Citizen Consultation

The concept of public participation was brought onto the agenda of urban design and planning prominently after the May events of 1968 [25]. Sherry R. Arnstein [26] was first to identify eight ways of participation: *manipulation, therapy, informing,* **consultation**, *placation, partnership, delegated power and citizen control*. After this study, it became more evident that facilitating participation practices do not necessarily grant design power to the citizens; they may manipulate them as well.

Following the Arnstein's ladder, the understanding of participation shifted towards greater democratization of the processes and deeper involvement of citizens [27][28][29]. This shift was in parallel with a theoretical shift or "the communicative turn" from rational planning to communicative and deliberative planning [30].

From the perspective of geospatial technologies, it is possible to track similar layers of transformation regarding the production and dissemination of geographic information. From top-down to bottom-up, referring to the public participation GIS (PPGIS), from "requested production" to "voluntary production" - geocrowdsourcing and finally, towards the wikification of GIS and Geoweb 2.0 technologies [7]. Preliminary examples of these kinds of initiatives are the "civic crowd" sponsored by the British Design Council, "Change by Us" by the cities of New York and Philadelphia, "Spacehive" by multiple actors in London, "Fix My Street", "Neighborland", "SeeClickFix", "Openplans" which are used for the collection of the ideas from citizens. These served as a motivation for us to initiate a field study to test Geoweb 2.0 in a real-life practice.

*Field Study 2: Citizen Consultation in the Planning of the Parks of Brussels*

This initiative was taken in January 2012 together with the Green Belgium Non-profit Organization which manages an educational network of 20.000 youngsters which are members of environment related "clubs".

The research process involved making a basic survey to identify the needs and aims of the organization. This was followed more than ten focus group meetings and email dialogues through which the scope and content of the user contributions were identified and various interface designs were realized. The usability of these interfaces was first tested with a limited number of users which represented relevant parties. After several improvements (including the French translation) the full version was made available to all parties in mid June, 2012.

In this study, Platform P2 was employed and customized as an instrument of dialogue between the youth movement of Brussels and green area managers. Establishing a dialogue was essential because of the age and power differences between the related parties.

In this setup, youngsters in Brussels were invited to represent their opinions and ideas using maps (geotags and polygonal zones), images and text. In parallel, the decision makers, park wardens and gardeners expressed their ideas and the problems they faced in a similar format. These two participant groups monitored what others think and wrote their own reviews. The scope and content of the user contributions were focused on specific aspects: *dreams, favorite places* and *problems*. All types of content were aggregated and overlaid together on the main page (Figure 5).

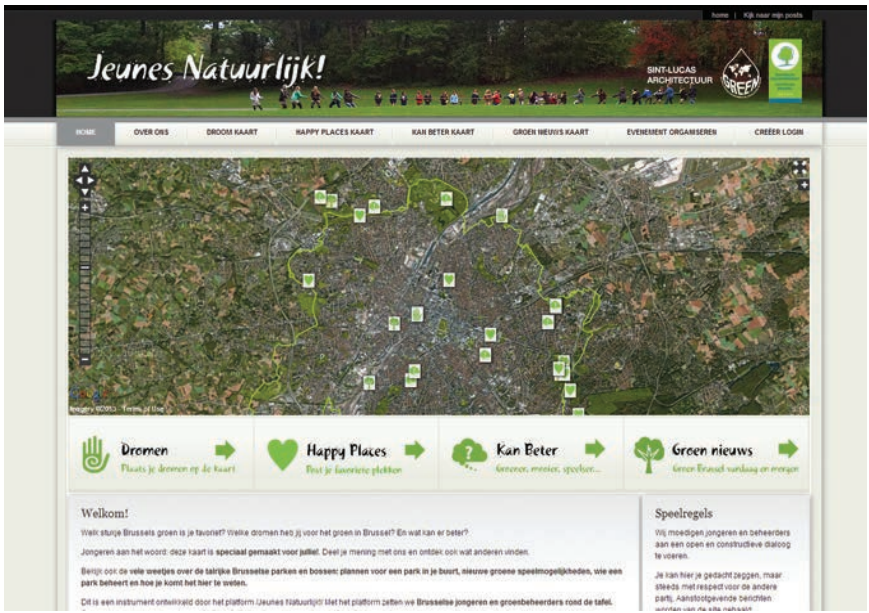

◀ Figure 5. Green Belgium Jeunes Natuurlijk! study.

On the map, individual categories were represented as individual layers with specific icons and clustered when needed to promote ease of use. The anonymisation of the user contributions helped the protection of their privacy and avoided possible disempowerment of the participants.

As a result of this study, we have learned that Platform P2 *can afford* the dynamical generation of collective thematic maps such as "dream maps", "favorite maps" and "problem/improvement maps". Through an import/export flow mechanism and advanced interaction tools provided with the openlayers library, the users were able to learn how to use the platform *intuitively* and share geolocated content without the need for a tutorial or a crash course.

An important quality was the multi-lingual nature of the contributions, which were intentionally harvested together to encourage the communication between French-speaking and Dutch-speaking youngsters as well as managers. These were perceived by the Green Belgium organization as a preliminary basis for establishing a sustainable reflective dialogue between youngsters and managers.

However, this dialogue was not balanced: managers contributed to very few posts (less than 15 percent), resulting in a biased conversation.

A major challenge was the fact that the goal was chosen by a non-profit organization without the participation of the actual users. Consequently, a number of on-site activities and promotional actions were necessary to motivate the users to sign up and provide feedback.

Another challenge was the lack of a legal framework to assign responsibility to the managers in the participation process. For this reason,

it was not possible to guarantee the inclusion of the constructed body of knowledge into decision making processes. This factor reduced the level of trust among the participants. In this context, in Arnstein's ladder of participation [26], this case can be considered as limited *consultation*.

### 2.3. Affordance 3: Urban Design Empowerment

Introduced in the previous section, the concept of *participation* assumes a power relation between the participant and a process to be participated into. Therefore, this concept is not sufficient to explain specific practices in which the citizens or organizations *independently* develop their own plans and projects.

In contrast to participation, *empowerment* can be considered as a better lens for describing enabling actions that gives authority.

In this context, based on Arnstein's ladder and focusing on the capacity of the visualization media for design empowerment, Senbel and Church [31] proposed six "instances" of design empowerment: *information, inspiration, ideation, inclusion, integration and independence*. In this theoretical frame, the highest level of empowerment is achieved when residents gain the capacity to create their own plans and visions and thus reach autonomy. From this perspective, Geoweb 2.0 platforms can be considered as a potentially affording design-empowering mechanism. Through technologies such as WikiGIS, alternative plans can be created by the public in an asynchronous and distributed manner to represent abstract forces shaping urban life; urban dynamics which are not usually accessible to designers and planning authorities [32]. Motivated by the above, a field study was specifically initiated for testing the potentials of platform P1 for design empowerment.

*Field Study 3: Urban Design Empowerment- the Independent Green Networks Plan*

In this study, the Brussels Environment Council (BRAL), a non-profit activist organization used the platform P1 to develop an extended and alternative version of the Brussels Green Network Plan. The platform was customized to fit in the needs of the BRAL team, consisting of experts with only a little knowledge of Geographical Information Systems. The requirements were analyzed through focus group meetings. These included the possibility to observe and draw over the previously created plans, such as the land use plan, the biological evaluation map and the older green network plan organized as separate base layers, as well as combining their new plans with these older plans. The open-source platform MediaWiki and the Google Mapper extension – in their original form – did not include this functionality so we had to develop custom applications and modify the extension to enable layering and create an "input-output" flow mechanism. In the modified version, when a user created a map it was possible to visualize it on any page using the import and export workflow (Figure 6). This system operated as a geo-RSS feed engine.

The research and development process involved usability analysis with limited amount of participants (n=6). We followed the diagnostic usability evaluation method to identify problems and gain an understanding of the difficulties that users face. The quality measures and metrics used in this study were unassisted task effectiveness, number of user errors, system errors and assists. We asked from the participants to perform 14 tasks that represent basic interactions related basic tasks like login and search, retrieve a topic, edit and format it, add a map to the discussion, add a placemark and mark an area on the map, place multiple maps on top of each other and logout. The feedback and results of this study were used to improve the efficiency, effectiveness of the platform and the satisfaction of the users.

▼ Figure 6. BRAL Green networks application.

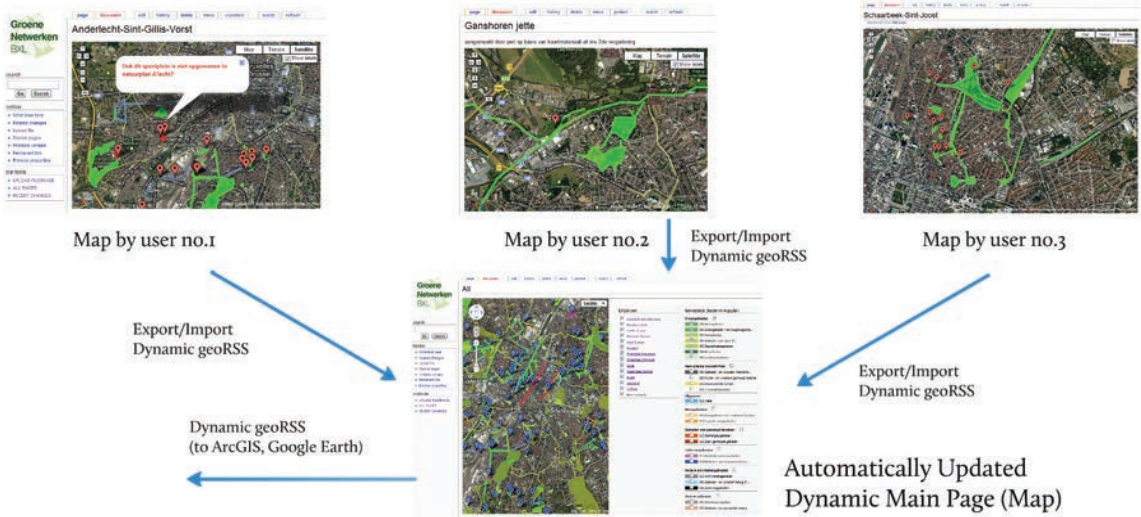

After updating the platform, the participants used it for three months in a collaborative manner and created a collective plan (Figure 7). An interesting aspect of this study was the inclusion of crowdsourced information into the design process. Specific maps created by citizen gardeners were made available and used as a basis to discuss the planning of future green networks. Together with the proposed plan, these were treated as a part of creative commons.

Two alternative views of the study were exported in both ArcGIS vector and hi-resolution raster image formats. At the end of May 2011, the final plan was presented and handed over to the Environmental Management Institute (IBGE) Study Office which was responsible for the preparation of the green networks section of the sustainable regional development plan.

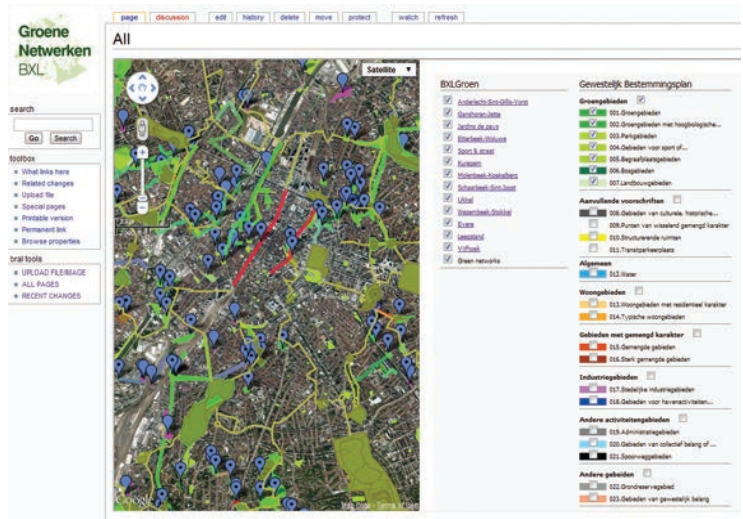

► Figure 7. The Alternative Green Networks Plan covering the green land use plan and crowdsourced data.

In this context, the knowledge that has been created through Platform P1 was officially transferred to planning authorities. In this sense, this case can be considered as a simple "independent design" example (level 6) in the empowerment scale of Senbel and Church (2011).

From this study we have learned that Geoweb 2.0 platform P1 can *afford* the distributed design of alternative plans as well as the inclusion of crowdsourced data. It was possible to export and import the final plans, organize them into layers by experts without any background knowledge on GIS software. The participants were highly committed and motivated because they independently chose the focus of their projects.

## 2.4. Affordance 4: Learning/Design Studio 2.0

The educational setting of the architectural schools depends on highly reflective practices which focus on the "design studio" as their central component. The design studio is a place where students learn experientially by designing their own projects through periodical critiques and collective reviews. During the critiques, they "seek to dwell in the moves of" an experienced designer (teacher) and "try to understand it by observation, imitation and picking out the essential features of the action" [33], in other words, they "know-in-action" [34]. The students are expected to consider their design alternatives together with the existing social and spatial urban environment and build relationships between these while redefining them. In this context, Geoweb 2.0 technologies provide various opportunities for the mediation of the dialogue between the design students and studio teachers. This dialogue can be brought to a level where a web 2.0 platform supports, augments and enriches the reflective learning processes. We propose to call this new setting "Design Studio 2.0" [8]. Design Studio 2.0 can potentially provide opportunities which are not fully available in a conventional design studio setting.

## Field Study 4: Design Studio 2.0

Two design studios have been organized at the KU Leuven Faculty of Architecture Campus Sint-Lucas Brussels to test the potentials of Geoweb 2.0 in design learning. The first one took place during the Spring Semester of 2012 with the participation of 34 international master's students from eleven different countries in Europe. They were motivated to explore the Luxembourg City center in groups and use their findings as a source of inspiration for the creating a new project for the design site (Figure 8). The design task was to understand what makes ordinary places in the city and how the locals relate themselves through their own human situations, events, meanings, and experiences.

▼ Figure 8. Geoweb 2.0 platform (P2) used in the first design studio, Spring 2013.

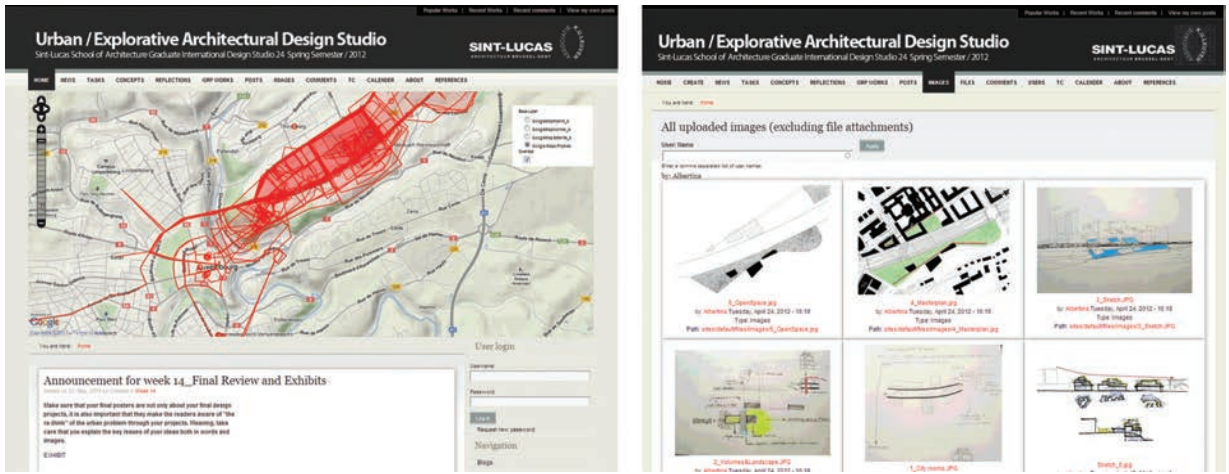

The second design studio took place during the Spring Semester of 2013. 27 international students from ten different countries in Europe participated in this course. Similar to the first studio, the students worked in groups and made a thorough urban analysis of the project site in a collaborative manner and shared these on the provided platform in the form of responses to the challenges issued by the studio coordinators. Using the constructed body of knowledge, they explored alternative ways of reconfiguring the spatial landscape of the focus area (Figure 9 on the next page).

Both of the two design studios were a result of an interactive inquiry process in which the first author was an active participant in the design studio as a teacher. In this sense, the knowledge generation in the two cases below involved "a continuing reflection on practice under real-time conditions"[35]. The experiences acquired from the first experimental

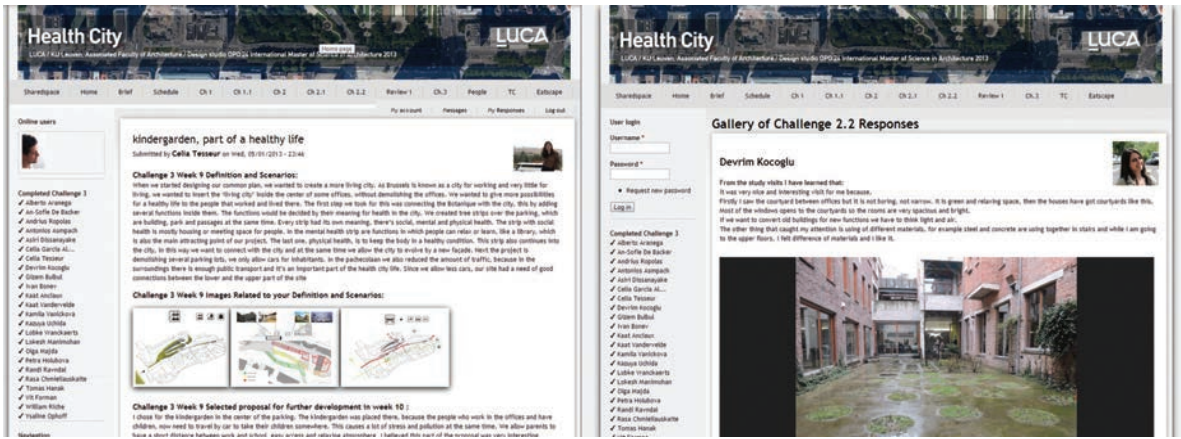

▲ Figure 9. The interface of the Platform used in the second studio, Spring 2013.

studio were reflected as improvements and readjustments onto the design of the web platform employed in the second studio.

We relied on multiple data sources to seek answers regarding the affordances of the Geoweb 2.0 platform. Among those were use logs collected through the proposed platform, student grades, an online questionnaire and a feedback meeting for each design studio. In addition to these, the first author took notes of the topics that arose during the process. The data was processed and sources were cross-compared while searching for the possible reasons for similar and conflicting observations [23].

In both of the studios, the Geoweb 2.0 platform P2 enabled us to extend the learning that took place in the design studio beyond the studio hours, to represent the design information in novel ways and allocate multiple communication forms.

Through the introduced setup, we were able to extend the reflective conversation in the design studio. Using collaborative mapping functionality, the students were able to collectively construct a shared memory of urban spaces which reportedly helped them to develop a better understanding of their project site. They were able to learn from their peers as well as the external experts.

Moreover, it was possible to combine conventional and online learning activities. By this way, the focus of the design studio was oriented more towards the students and the learning processes. The teacher-student relationship made it easier for the participants to trust the Geoweb platform as a participation tool.

The students frequently commented on each other's works and constructed a common understanding.

In addition to the above, through the analysis of the use logs we found that the students' participation in the introduced web platform was positively related to their progress up to a certain point [8][23].

### 2.5. Affordance 5: Urban Design Research

Geoweb 2.0 platforms are also potential tools for learning how specific spatial qualities are perceived by the inhabitants. They can provide alternative interfaces for web and mobile browsers, enabling the input of ratings as well as output in the forms of maps and dashboards [23]. By using Geoweb 2.0 platforms, dynamic knowledge acquired through lived experience can be used as a vital resource for research and design purposes. Alternative location-based maps can be created by involving the public to represent urban dynamics. Multiple perspectives of individuals, social groups and organizations can be dynamically represented and socially discussed. By working with alternative depictions of urban environments, one can simultaneously account for representations of the existing context and imaginations of different realities.

*Field Study 5: Walkability*

In order to examine the potentials of the Geoweb 2.0 platform a research study was conducted in Brussels. It aimed at the collection of experiential information from the inhabitants of a specific neighborhood. A triangular path surrounding a major Square was chosen as a test zone. This area is one of the most controversial and segregated places in the city, which happens to include the North Station and an ethnic shopping street. Six participants were asked to walk around the neighborhood and continuously express their opinions on the walkability problems of the location. The first author accompanied and interviewed each participant during a two-hour walk-along, while making location-based notes and taking photos.

The collected information was uploaded to the platform via:
- A mobile device / geolocated notes and photos, *during the walk, on location*
- A desktop browser, *after the walk, based on the notes*

After the participatory study, a joint heat map was constructed using the walkability ratings of the six contributors (Figure 10). This heat map renders a predefined gradient based on the intensity of a data point. The more negative perception of the participants, the more it shifts towards red. The heat map enabled the dynamic visualization of three dimensional data, in which two dimensions represented Cartesian coordinates and the third dimension was used for visualizing the intensity of walkability as a datapoint in relative comparison to the absolute maximum of the dataset. Using the datapoints, an alpha map was created using a radial gradient with 0.1 alpha as the maximum value which fades out to alpha=0. Then these values are converted to RGB. This method gave us the flexibility to build a customized color shift from alpha 255 to 0 and control the radius of the data points. The intensity was shown as a color; red (hot) for the maxima and blue (cold) for minima.

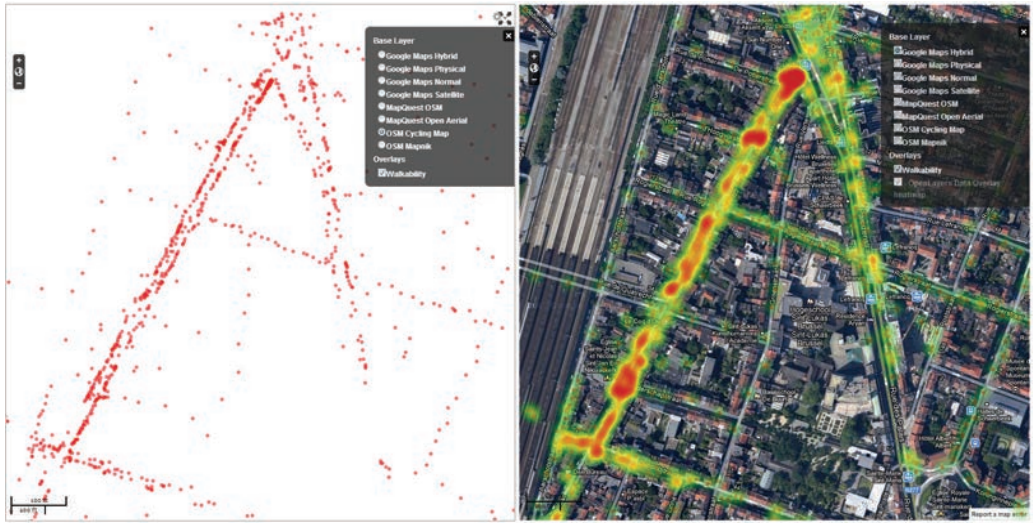

◄ Figure 10. Screenshot from the Geoweb 2.0 platform: a heat map (on the right) generated by the more than 300 walkability problem points (on the left).

In this study the Geoweb 2.0 platform enabled us to extract and diagnose a significant number of problems. By using the heat maps in combination with the walk-along experiences, we were able to develop ideas for solutions to design and planning problems which may provide measurable benefits to the inhabitants. For instance, we have suggested limited pedestrianization of the shopping street with a high number of negative walkability ratings (red on the heat map on Figure 10). We recommended the redesign of the pedestrian crossings at two points, again highlighted as red on the map and referenced during the interviews by all of the participants. Besides the identification of problematic areas in the pilot study area, by analyzing the location-based notes entered on the content management platform, it was possible to extract various interrelated spatial qualities [24]. Following these, the results were anonymized and published as a paper with the consent of the participants.

## 3. CONCLUSIONS, SUCCESS FACTORS AND RECOMMENDATIONS

In this paper, reflecting on our findings from real-life practices, we have discussed five major action possibilities (affordances) of two Geoweb 2.0 platforms for the deliberation and participatory design of alternative urban projects. These affordances were related to specific users (Table 1).

| Affordance | Geoweb 2.0 Platform | Users | Affordance Level |
|---|---|---|---|
| Conversations on Alternative Projects | P1 | Civil Society, Planners and Development Agencies | Feedback on the Geolocated Representations of Alternative Urban Projects |
| Citizen Consultation | P2 | Non-profit Environmental Organizations and Managers | Exchanging Ideas, Learning Preferences and Problem Consultation |
| Urban Design Empowerment | P1 | Non-profit Planning Organization | Analysis, Mapping and Alternative Plan Development |
| Learning /Design Studio 2.0 | P2 | Urban Design Students, Teachers, Experts, and Inhabitants | Collective Analysis, Mapping, Exchanging Design Ideas, Learning from the Peers and Experts |
| Design Research | P2 | Urban Design researchers, Inhabitants and Planners | Ratings of Urban Qualities, Heat Maps, Learning from the Inhabitants |

◄ Table 1. An overview of the affordances, users and affordance levels.

As a result of the first field study (Section 2.1), we have learned that geolocation can serve as an integration tool for representation and discussion of the projects. Using the proposed platform P1 it was possible to divide a large body of text into individual place-based strategies and geolocate them on a map as interrelated place marks with explanations.

In the second field study (Section 2.2), we observed that Platform P2 *afforded* the dynamical generation of collective thematic maps. Through the use of this platform, ordinary citizens were able to share geolocated content intuitively, without the need for a tutorial or a crash course.

In the third field study (Section 2.3), Geoweb 2.0 platform P1 enabled the distributed design and dynamic representation of alternative plans by non-designers who were a member of a non-profit organization. They were able to import existing plans and crowdsourced data; organize them into layers and create their own plans without any background knowledge on GIS software.

During the fourth study on learning (Section 2.4) the Geoweb 2.0 platform P2 enabled us to extend the learning that took place in the design studio beyond the studio hours, to represent the design information in novel ways and allocate multiple communication forms. Through the analysis of the use logs, we found that the students' participation in the introduced web platform was positively related to their progress up to a certain point.

In the fifth field study (Section 2.5), Geoweb 2.0 platform empowered us to research and map a significant number of problems in a specific neighborhood. By using heat maps we were able to develop novel design addressing these issues and extract various interrelated spatial qualities.

*Success factors and recommendations for best practices*

In the field studies reviewed in the previous section, we have observed several common and interrelated success factors. Among those the most important were (1) **the goals and goal setting mechanisms,** (2) **equality,** (3) **authorship,** (4) **privacy** and (5) **trust**:

**Goals:** In cases when *the goals* were set and shared in a participatory manner, the motivation level of the participants was higher. For instance, the users who contributed to the third field study (2.3) (members of a non-profit organization) independently chose the focus of their projects and therefore they were highly committed. In contrast, in the second field study (2.2), the goal was chosen without the participation of the actual users. Consequently, a number of on-site activities and promotional actions were necessary to motivate the users to sign up and provide feedback. These findings also support ACCOLADE Project [36] which stressed that the construction of shared understanding of goals is essential for better collaboration practices. In this sense it is essential to facilitate participation in the early stages of design, including the goal setting and sharing phases

**Equality**: Large-scale online participation (*crowdsourcing*) relies on the hypothesis that if the contributor pool is big enough, it can represent the participants. However several factors such as familiarity with web technologies, access to internet and age of the participants affect their engagement frequency and thus *equality* in representation. For instance in the field study (2.2) the managers contributed to very few posts, resulting in a biased conversation and affecting the process negatively. Therefore for the future practices promoting equality in representation can be a valuable strategy for increasing the engagement levels of the participants.

**Authorship:** Geoweb 2.0 enabled participation and design raises several questions over the *authorship* and intellectual property rights regarding the design products. The international legal frameworks and regulations regarding these practices are still underdeveloped and this problem negatively affects their reliability and validity. In all of the cases reported in the previous section, we tried to address this challenge by clearly informing the participants about the intellectual ownership structure and their rights on the final product before the initiation of the participation process. In two of our cases (2.2 and 2.3) the participants signed an online crowdsourced contributor agreement and the final product was licensed as a part of creative commons by the organizing non-profit institutions. The walkability analysis case (2.5) was a research study. The results were published as a paper with the consent of the participants. In the design studio learning case (2.4), the students kept the ownership of their individual designs according to the legal education framework of the university and only agreed to distribute images of their works online.

**Privacy:** Geoweb 2.0-enabled participation practices naturally involve the collection of data from the participants. But this data is not neutral; it contains traces of *private* information. It is of utmost importance to treat this information sensibly and carefully. The participants should be informed clearly on how their digital activities are being collected and processed. In all of studies reported above we tried to accomplish this through brief introduction sessions. Moreover, when necessary, the anonymization of the user contributions helped the protection of their privacy and avoided possible disempowerment of the participants.

**Trust:** Finally, *trust* was a key factor which affected the participation processes. The perceived trustworthiness of the participation processes and the technological tools was closely tied to the goal-setting, representation equality, the treatment of authorship issues as well as privacy matters. In the fourth field study (2.4), the teacher-student relationship made it easier for the participants to trust the Geoweb platform as a participation tool. On the other hand, in the second field study (2.2), it was difficult to establish trust due to the status and age differences between the stakeholders.

In conclusion, the cases presented in our paper can be seen as field studies on giving a voice to non-governmental organizations, acceptance of user created data as a valid resource and its inclusion in the planning practices. The proposed Geoweb 2.0 platforms successfully supported the subsequent communicative and participatory processes and the initial outcomes conformed to the intentions of our studies. However, affording participation did not automatically guarantee the empowerment of the citizens. The ultimate success indicator for similar future applications will be the extent to which the plans and messages of the participants are taken on board by the authorities.

A possible future direction for this research is to develop a sustainable monitoring system that joins the scattered knowledge on rapidly changing social needs in various neighborhood infrastructures. This system can include a mobile platform for gaining insight into the various qualities of specific urban neighborhoods. This would be possible through the creation of alternative representations that support the multi-scalar and trans-territorial planning and design which is potentially beneficial for the inhabitants, planners and governing authorities.

Another idea for the future is to integrate Geoweb 2.0 applications into cutting edge sustainability research to address the pressing environmental challenges such as the climate change, air pollution and loss of biodiversity in the planning processes. In this sense, a possible solution would be to establish an extended platform and reframe the participatory planning practices as a research network composed of multidisciplinary teams of researchers, educators, managers, policymakers and other relevant stakeholders.

Furthermore, combination of user created geographical information and experiential feedback with transdisciplinary computational models such as UrbanSim [37] will also be potentially beneficial from the perspectives of urban design and planning as well as systems research.

## ACKNOWLEDGEMENTS

This research was supported by a three-year type (B) "Prospective Research for Brussels" postdoctoral research grant from the Brussels Capital Regional Government, Institute for the Encouragement of Scientific Research (INNOVIRIS) awarded to Burak Pak, promoted by Johan Verbeke.

**Burak Pak and Johan Verbeke**

KU Leuven Faculty of Architecture
Campus Sint-Lucas Brussels
Paleizenstraat 65-67, 1030 Brussels, Belgium

burak.pak@kuleuven.be, johan.verbeke@kuleuven.be